\documentclass[aps,prl,showpacs,twocolumn,amsmath,amssymb,superscriptaddress]{revtex4}
\usepackage{graphicx}
\usepackage{Jonasmacros}
\usepackage{bm}

\newcommand{\br}{{\bf r}}

\newcommand{\beqa}{\begin{eqnarray}}
\newcommand{\eeqa}{\end{eqnarray}}

\begin{document}
\title{Detection and Cloaking of Molecular Objects in Coherent Nanostructures Using Inelastic Electron Tunneling Spectroscopy}
\author{J. Fransson}
\email{jonas.fransson@fysik.uu.se}
\affiliation{Department of Physics and Astronomy, Uppsala University, Box 534, SE-751 21\ \ Uppsala, Sweden}

\author{H. C. Manoharan}
\affiliation{Department of Physics, Stanford University, Stanford, CA 94305, USA}
\affiliation{Stanford Institute for Materials and Energy Sciences, SLAC National Accelerator Laboratory, Menlo Park, California 94025}

\author{A. V. Balatsky}
\affiliation{Theoretical Division, Los Alamos National Laboratory,
Los Alamos, NM 87545, USA}
\affiliation{Center for Integrated Nanotechnologies, Los Alamos National Laboratory,
Los Alamos, NM 87545, USA}
\date{\today}

\begin{abstract}
We address quantum invisibility in the context of electronics in nanoscale quantum structures. We make use of the freedom of design that quantum corrals provide and show that quantum mechanical objects can be hidden inside the corral, with respect to inelastic electron scattering spectroscopy in combination with scanning tunneling microscopy, and we propose a design strategy. A simple illustration of the invisibility is given in terms of an elliptic quantum corral containing a molecule, with a local vibrational mode, at one of the foci. Our work has implications to quantum information technology and presents new tools for nonlocal quantum detection and distinguishing between different molecules. 
\end{abstract}
\pacs{IETS, quantum invisibility, quantum corral, STM, nanoscale engineering}
\maketitle

As we approach the quantum limit for a wide range of experiments and technologies, it is rather natural to ask whether it is possible to hide information from the measurements. A recent theoretical study suggested that a cloak of invisibility is in principle possible in optical applications, using the freedom of design offered by metamaterials in order to redirect electromagnetic fields \cite{pendry2006}. It has since been shown that a copper cylinder may be hidden inside a metamaterial cloak constructed according to the theoretical prescription \cite{schurig2006}.

Physical and chemical  signal detection is based on the interactions between the probe and the sample. Those interactions are probed within a certain frequency range, which itself depends on the specific property we are probing, e.g spin or charge response. Thus, by eliminating or weakening the effect of the relevant interactions within the operational frequency range, the objects can be made invisible from the detector. In this sense, one has been able to show that metamaterials can be used as a cloaking device. One example is a copper cylinder, that was made \emph{invisible} in the microwave frequency band range. Here, we discuss detection and invisibility within the THz band. The discussion is based on the ability to intentionally engineer the properties of quantum structures with tailored  properties in the THz band range. This high frequency regime is accessible for nano structures and it opens the window for detection of species down to single molecules and atoms, based on their chemical (vibrational) signatures.

In this Letter, we address invisibility in the context of electronics in nanoscale quantum structures, for which we use quantum corrals as a prototype coherent device. We take advantage of the freedom of design that quantum corrals provide and show that quantum mechanical objects located inside the corral can be made invisible with respect to certain measurement. We first need to ask how specific species of quantum matter can be detected. This can be accomplished through chemical identity of atoms and molecules for which molecular vibrations function as a \emph{fingerprint} of species identity. In this case, the relevant frequency scale is THz rather than MHz. Remarkably, scanning tunneling microscopy (STM) can provide access to these frequency scales through the inelastic electron tunneling spectroscopy (IETS) process, and can simultaneously give access to the challenging length scales necessary to probe and manipulate individual quanta of matter and excitations.  We propose a method for detecting and manipulating quantum invisibility based on THz cloaking of molecular identity in coherent nanostructures.

Using IETS combined with STM we thus have an experimental tool that enables identification of molecules through detection of certain \emph{fingerprint frequencies} which correspond to vibrational modes of the molecules. Recently, CO molecules were used on Cu(111) surfaces for molecular assembly and quantum corrals \cite{moon2008} and well-known molecular excitations exist for this materials system. Quantum corrals with elliptic geometries have recently been employed for discussion of mirage effects \cite{manoharan2000,aligia2001,hallberg2002,gadzuk2003,correa2005,rossi2006,moonNP2008}, where a magnetic adsorbate atom or molecule is located in one focus of the ellipse while the (STM) measurement is performed nonlocally in the second, empty, focus. 

To be specific, we consider an elliptic quantum corral containing a single molecule with a local vibrational mode, located at one of the foci; we argue that the vibrational mode should be measurable at the second, unoccupied, focus through IETS combined with STM. Theoretically, it has been shown that vibrational modes in a molecular structure adsorbed on a metallic surface would be measurable by means of IETS in a narrow energy range around the vibrational mode \cite{balatsky2006,franssonIETSfriedel2007}. Moreover, the inelastic scattering generates Friedel oscillations in the local surface density of states (DOS), thus enabling spatial imaging of the inelastic signatures. In the present context, the standing waves generated inside the corral by the inelastic scattering at the occupied focus are reflected and reassembled at the unoccupied focus, thus creating a finite ghost response. Within the same framework and exploiting engineering abilities of the quantum corral for another purpose, we here suggest that the molecule can be made undetectable.


As an experimentally realizable system \cite{moon2008}, we consider elliptical quantum corrals comprised of $N$ CO molecules on a Cu surface. The CO molecules generate a vibrational mode $\omega_0\simeq4$ meV caused by translational frustration when placed on a metallic, e.g. Cu, surface. The presence of the quantum corral generates a modified LDOS in a neighborhood around itself. Using the scattering theory proposed by Fiete and Heller \cite{rodberg1967,heller2003}, we calculate the modified LDOS built up by the quantum corral. Our calculations indicate that the IETS contribution from molecules on the walls is small with respect to the signal of the focal molecule for these geometries. Thus, the contribution from the vibrations in the wall does not significantly contribute to the effect we are studying, but merely provides a small modification to the overall background. Therefore, we neglect the inelastic contribution from the wall and, for simplicity, use the elastic scattering theory in the construction of the LDOS generated by the molecules in the wall of the quantum corral. 

We employ a $T$-matrix approach \cite{rodberg1967,heller2003} to generate the electronic structure imposed on the surface by the elliptical quantum corral. The corral is generated by placing molecules equidistant along the ellipse $(x/a)^2+(y/b)^2=R^2$. The two foci in the ellipse are located at $\bfr_\pm=(\pm c,0)$ where $c=\sqrt{a^2-b^2}$. Having established the electronic structure of the quantum corral, we proceed by including the inelastic effects from a molecular impurity located at the \emph{molecular focus} $\bfr_+$ (the second focus at $\bfr_-$ is henceforth referred to as the \emph{empty focus}).

All vibrational modes have the fingerprint frequency $\omega_0$ and are assumed to be the same as they originate from the same type of molecules. We use the Hamiltonian for the local vibrational modes, coupled to electrons via Holstein coupling \cite{holstein1959} with interactions assumed to occur only at the single impurity site, so that
\begin{align}
\Hamil=\sum_{\bfk\sigma}\dote{\bfk}\cdagger{\bfk}\cc{\bfk}
    +\omega_0\bdagger{}\bc{}
    +\lambda\sum_{\bfk\bfk'\sigma}\csdagger{\bfk\sigma}\cs{\bfk'\sigma}(\bdagger{}+\bc{}).
\label{eq-Ham}
\end{align}
Here, a surface electron is created (annihilated) by $\cdagger{\bfk}\ (\cc{\bfk})$ at the energy $\dote{\bfk}$. The strength of the electron-vibron interaction is given by the parameter $\lambda$, whereas $\omega_0$ is the mode of the bare vibron which is created (annihilated) by $\bdagger{}\ (\bc{})$.

\begin{figure}[t]
\begin{center}
\includegraphics[width=8.5cm]{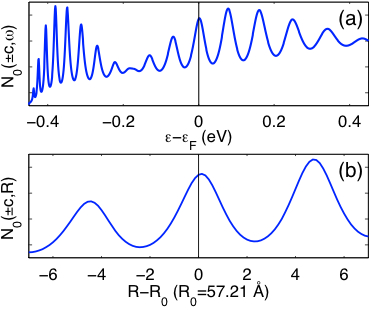}
\end{center}
\caption{a) Typical electronic structure of the quantum corral at the ellipse foci $\pm c$. The underlying ellipse is given by $(x/a)^2+(y/b)^2=R^2$ with $R=57.21$ \AA, and $a/b=1.5$, comprised of 40 molecules. b) Local DOS at the ellipse foci at the Fermi level as function of the ellipse radius $R$. The electronic structures were calculated using quadratic energy dispersion $E_\text{surf}(k)-E_F=E_0+\hbar^2k^2/2m^*$, with $E_0\simeq-0.45$ eV for Cu(111), and $m^*=0.38m$, where $m$ is the free electron mass.}
\label{fig-qc1}
\end{figure}
The features we are considering are expected to be seen in the second derivative of the tunneling current with respect to the bias voltage $V$ in real space, i.e. $\partial^2I(\bfr,V)/\partial V^2$. This quantity is directly proportional to the frequency derivative of the local DOS. In second order perturbation theory (sufficient for weak electron-vibron coupling), this amounts to taking the frequency derivative of the correction to the density of states, $\delta N(\bfr,\omega)$, due to the influence of the impurity scattering.  The real space electron Green function (GF) is given by\begin{align}
G(\bfr,\bfr';\omega)=&G_0(\bfr,\bfr';\omega)
\nonumber\\&
    +G_0(\bfr,\bfr_+;\omega)\Sigma(\omega)G_0(\bfr_+,\bfr';\omega),
\label{eq-GF}
\end{align}
where $G_0(\bfr,\bfr')$ is the GF for electronic structure of the quantum corral without the molecular impurity \cite{franssonIETSQC2008}, whereas $\bfr_+$ is the position of the impurity inside the quantum corral. The self-energy $\Sigma$ is given by
\begin{align}
\Sigma(i\omega)=&
	\lambda^2\sum_\bfk
	\biggl[\frac{n_B(\omega_0)+f(\dote{\bfk})}{i\omega+\omega_0-\dote{\bfk}}
\nonumber\\&\hspace{2cm}
	+\frac{n_B(\omega_0)+1-f(\dote{\bfk})}{i\omega-\omega_0-\dote{\bfk}}\biggr],
\label{eq-S}
\end{align}
where $n_B(x)$ and $f(x)$ are the Bose and Fermi functions, respectively.

For later reference, we denote the unperturbed LDOS of the quantum corral by $N_0(\bfr,\omega)=-\im{G_0(\bfr,\bfr;\omega)}/\pi$ and the correction due to the electron-vibron coupling by
\begin{align}
\delta N(\bfr,\omega)=&
	-\frac{1}{\pi}\im{\{G_0(\bfr,\bfr_+;\omega)\Sigma(\omega)G_0(\bfr_+,\bfr;\omega)\}}.
\end{align}
The dressed LDOS $N(\bfr,\omega)$ is then given by $N(\bfr,\omega)=N_0(\bfr,\omega)+\delta N(\bfr,\omega)$. Generally, the differential conductance $\partial I(\bfr,V)/\partial V$ between the STM and the structure is directly proportional to the LDOS $N(\bfr,\omega)$, and the second bias voltage derivative $\partial ^2I(\bfr,V)/\partial V^2\propto\partial N(\bfr,\omega)/\partial\omega$. The latter derivative, $\partial ^2I(\bfr,V)/\partial V^2$, is used for inelastic tunneling spectroscopy (IETS) using STM. It is therefore sufficient to study the LDOS and its frequency derivative in order to account for the expected qualitative results in an STM measurement.

Typically, the electronic structure of the elliptic quantum corral is an oscillating function of the energy. Fig. \ref{fig-qc1} (a) illustrates a typical DOS at the two foci. The geometry in this particular case gives rise to a peak in the DOS at the Fermi level $\dote{F}$. Whether the local DOS at the foci peaks at the Fermi level is a geometrical matter of the structure. This is illustrated in Fig. \ref{fig-qc1} (b), which displays the local Fermi level DOS at the foci for varying ellipse radius $R$, and it is clear that the DOS at the Fermi level is an oscillating function of the radius as well. The width of the density peaks are related to the fact that the quantum corral is comprised of discrete atoms or molecules and is not an entirely closed structure, although the wall nevertheless create quantum confinement. The wave functions, thus built up within the confinement potential, have a non-negligible leakage to the surrounding environment and this generates a broadening of the confined states. The level broadening is accounted for in the employed $T$-matrix approach \cite{rodberg1967,heller2003}. The general conclusion drawn is that the quantum corral can be engineered, or designed, to meet certain requirements set by the desired functionality.

Quantum invisibility in our context would be the ability to hide information from detection, which in a broader sense implies information storage away from a perturbing environment. Hiding information away from reading technology would in principle mean an extremely small interaction with the surroundings and would, hence, allow for extremely long decoherence times. We might also think of other applications of this ability to make physical objects invisible from the measurements. For instance, we can create devices which can individually distinguish between different types of molecules, or other types of molecular or mesoscopic objects, by determining whether the quantum object carries a predefined fingerprint frequency.

In particular, the functionality we propose in this Letter is the ability to detect and/or hide molecules adsorbed at one focus inside the elliptic quantum corral, which have their fingerprint frequency within a certain frequency range. It is well known that perturbing the electronic structure in one focus generates a mirror response at the second focus, 
which presumably is empty \cite{manoharan2000,gadzuk2003}. However, the remote projection of an object at one focus is efficient only when the local DOS at the foci peaks, or is significant, at the particular energy of interest. The reason is straightforward: A large density allows for a large response while the response becomes suppressed for a small density. This rule thus enables a dual functionality of the device we propose, namely both detection within a certain frequency range and alternately quantum invisibility of molecular objects.

\begin{figure}[t]
\begin{center}
\includegraphics[width=8.5cm]{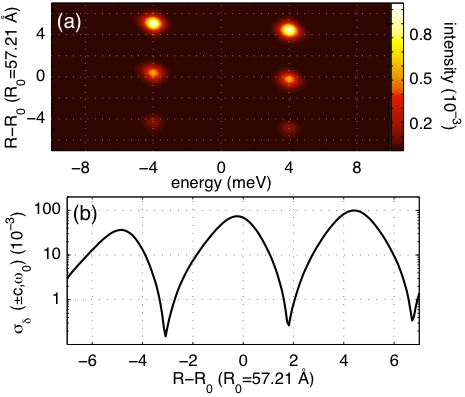}
\end{center}
\caption{(Color online) a) Size and energy dependence of the response $|\partial_\omega\delta N(\bfr_-,\omega)|^2$ at the empty focus $\bfr_-=(-c,0)$. b) Size dependence of the restricted deviation $\sigma_{\omega_0}$ at the empty focus. Here we use $R_0=57.21$ \AA, $\omega_0=4$ meV, at $T=4$ K. The restricted deviation has been calculated by integrating over the interval $(\omega_0-\delta,\omega_0+\delta)$, with $\delta=4$ meV.}
\label{fig-d2I}
\end{figure}

In order to quantify our assertions and to make contact with realistic experimental situations, consider the elliptic quantum corral corresponding to the local DOS at the foci given in Fig. \ref{fig-qc1} (a). From the above discussion we thus expect the response to the inelastic scattering to be an oscillating function of the ellipse radius. Moreover, following the argument in Refs. \cite{balatsky2006,franssonIETSfriedel2007} we expect sharp features in the response $\partial ^2I/\partial V^2\propto\partial\delta N(\bfr,\omega)/\partial\omega$ at the inelastic resonances $\omega=\pm\omega_0$ for low temperatures. These expected features of the response to the inelastic scattering are verified in Fig. \ref{fig-d2I} (a), where we plot the response $|\partial\delta N(\bfr,\omega)/\partial\omega|^2$ at the empty focus $\bfr_-=(-c,0)$ as a function of the ellipse radius and energy. It is readily seen that the response is an oscillating function of the radius and that it peaks around the fingerprint frequency $\pm\omega_0$.

We characterize the quality of the response from the inelastic scattering by defining the integrated deviation from the unperturbed response by
\begin{align}
\sigma(\bfr,R;\omega,\omega_0)=\sqrt{\int
	\biggl(\frac{\partial}{\partial\omega}\delta N(\bfr,\omega)\biggr)^2d\omega}.
\label{eq-dev}
\end{align}
A large value of this function clearly describes that the response to the inelastic scattering significantly deviates from the unperturbed response. Moreover, since the response $\partial\delta N(\bfr,\omega)/\partial\omega$ is expected to be large only in a narrow interval around the inelastic resonances $\pm\omega_0$ and small anywhere outside, this function provides a good measure of the deviation caused by the electron-vibron coupling. In the experimental situation, however, one would measure the response in a finite interval around the inelastic resonance, e.g. in the interval $(\omega_0-\delta,\omega_0+\delta)$, $\delta>0$. It therefore makes sense to restrict the integration in Eq. (\ref{eq-dev}) to this interval, and we call this function $\sigma_\delta(\bfr,\bfR;\omega,\omega_0)$. In Fig. \ref{fig-d2I} (b) we plot this restricted deviation at the empty focus $\bfr_-$ as function of the ellipse radius $R$, corresponding to the setup in Fig. \ref{fig-d2I} (a). This plot clearly illustrates the possibility to engineer the structure in order to maximize measurability of the electron-vibron response.

The ability to purposely engineer the quantum structure has wider implications than the above discussion about optimization of the response from the inelastic mode. For a given quantum structure, the underlying electronic structure is an oscillating function of the energy. As a such, one can use the presence or absence of underlying electronic density for the purpose of displaying or hiding the inelastic response from the impurity at the molecular focus.

\begin{figure}[t]
\begin{center}
\includegraphics[width=8.5cm]{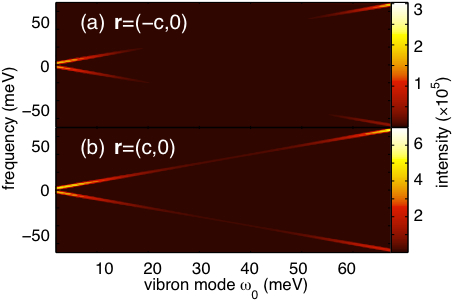}
\end{center}
\caption{(Color online) Dependence of the response $|\partial_\omega\delta N(\bfr,\omega)|^2$ on the vibron mode $\omega_0$ at a) the empty focus $\bfr=(-c,0)$, and b) the molecular focus $\bfr=(c,0)$, in a corral with a density peak in the foci at the Fermi level. Here we use $R_0=57.21$ \AA, at $T=4$ K. The two branches correspond to the two inelastic resonances at $\omega=\pm\omega_0$.}
\label{fig-Q1}
\end{figure}

\begin{figure}[t]
\begin{center}
\includegraphics[width=8.5cm]{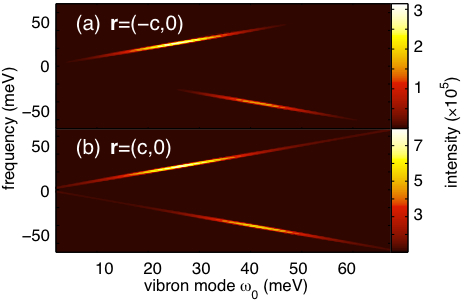}
\end{center}
\caption{(Color online) Same as in Fig. \ref{fig-Q1} for a corral with low electron density in the foci at the Fermi level. Here we use $R_0=60.21$ \AA.
}
\label{fig-Q2}
\end{figure}

Assume that the corral is optimized for a peaked electron density at the Fermi level in the foci. As seen in Fig. \ref{fig-d2I}, there is a detectable response at the empty focus whenever the vibrational mode of the impurity at the molecular focus lies within the width of the density peak. If, on the other hand, the vibrational mode fall outside the width of the density peak the response at the empty focus becomes less pronounced, and eventually invisible. A typical example of this discussion is illustrated in Fig. \ref{fig-Q1}, which shows the inelastic response at (a) the empty, and (b) molecular focus for varying vibrational modes of the impurity at the molecular focus. The two branches correspond to the inelastic resonances at $\pm\omega_0$. The plot clearly illustrates that the response in the empty focus decays as the vibrational mode starts to fall outside the width of the density peak, and eventually for sufficiently large vibrational modes the impurity becomes invisible to the probe at the empty focus. The returning signal for even larger modes arise because of the presence of another density peak at larger energies. All these features occur in the empty focus although the inelastic response at the molecular focus is sufficiently strong for detection. By the same token, we can design the corral to have a low electron density at the Fermi level in the foci. In this situation, impurities with low vibrational modes become invisible to the probe at the empty focus (see Fig. \ref{fig-Q2}). The asymmetry in the response with respect to resonance frequency is due to the fact that the underlying electron density is not symmetric around the Fermi level for the given geometry.

In summary, we have considered the engineering of quantum structures for precisely designing functionality.  By way of example, using elliptic quantum corrals with radii between 50 and 65 \AA, comprising about 40 atoms or molecules, we show that the geometry of elliptical quantum corrals can be optimized for large or small inelastic response at one focus, generated by the vibrational mode or \emph{fingerprint frequency} of an impurity at the other focus. Such optimization opens possibilities for detection devices that can be used to distinguish between different species of molecules. Moreover, functional design of the quantum structures enables cloaking, or quantum invisibility, of quantum objects.

The authors thank I. Grigorenko for useful discussions. This work has been supported by US DOE, LDRD and BES, and was carried out under the auspices of the NNSA of the US DOE at LANL under Contract No. DE-AC52-06NA25396 (J.F. and A.V.B), by the US DOE at SLAC under Contract No. DE-AC02-76SF00515 (H.C.M), and by the NSF (H.C.M). J.F. thanks Swedish Research Council (VR) for support.

\end{document}